\documentclass[prb,twocolumn,amsmath,amssymb]{revtex4}
\usepackage{natbib}
\usepackage{graphicx}         
\usepackage{dcolumn}         
\usepackage{bm}                 

\begin{document}

\preprint{}

\title{Spin precession and modulation in ballistic cylindrical nanowires
due to the Rashba effect}

\author{A. Bringer}
\affiliation{Peter Gr\"unberg Institute (PGI-1) and
JARA-Fundamentals of Future Information Technology,
Forschungszentrum J\"ulich GmbH, 52425 J\"ulich, Germany}

\author{Th. Sch\"apers}
\email{th.schaepers@fz-juelich.de} \affiliation{Peter Gr\"unberg
Institute (PGI-9) and JARA-Fundamentals of Future Information
Technology, Forschungszentrum J\"ulich GmbH, 52425 J\"ulich,
Germany}

\date{\today}

\begin{abstract}
The spin precession in a cylindrical semiconductor nanowire due to
Rashba spin-orbit coupling has been investigated theoretically
using an InAs nanowire containing a surface two-dimensional
electron gas as a model. The eigenstates, energy-momentum
dispersion, and the energy-magnetic field dispersion relation are
determined by solving the Schr\"odinger equation in a cylindrical
symmetry. The combination of states with the same total angular
momentum but opposite spin orientation results in a periodic
modulation of the axial spin component along the axis of the wire.
Spin-precession about the wires axis is achieved by interference
of two states with different total angular momentum. Because a
superposition state with exact opposite spin precession exists at
zero magnetic field, an oscillation of the spin orientation can be
obtained. If an axially oriented magnetic field is applied, the
spin gains an additional precessing component.
\end{abstract}

\maketitle

\section{\label{sec:intro}Introduction}

Semiconductor nanowires are almost ideal objects for studying quantum
effects and electron interference phenomena. The use of the bottom-up
approach for nanowire growth simplifies the preparation substantially
and allows us to create novel confinement schemes, such as axial or
radial heterostructures.\cite{Thelander06,Lu06} The large
surface-to-volume ratio of nanowires means that surface properties
are crucial for discussions of transport properties, so that low
band-gap semiconductors, e.g. InAs, InN, or InSb, are particularly
interesting. In these systems, the Fermi level at the surface is pinned
inside the conduction band,\cite{Lueth10a} and an accumulation layer is
formed. This guarantees that the conductance is sufficiently large even
at low nanowire radius. The presence of the surface accumulation
layer means that a tubular conducting channel is formed, and this shape
of the conductor has important implications for the magnetoconductance
of the nanowires. An example is the theoretical prediction and
experimental confirmation of flux-periodic oscillations in nanowires
with a magnetic field applied along the wire axis.
\cite{Tserkovnyak06,Richter08} The electronic states of a cylindrical
two-dimensional electron gas in a transverse magnetic field were
calculated by Ferrari \emph{et al.},\cite{Ferrari08,Ferrari08a} while
Magarill \emph{et al.} \cite{Magarill96,Magarill98} discussed the
kinetics of electrons in a tubular conductor.

Many concepts have been developed for planar semiconductor layer
systems that make use of the spin degree of freedom for device
structures. The best-known example is the spin field-effect
transistor,\cite{Datta90,Egues03,Schliemann03} which uses the
gate-controlled spin-precession induced by the Rashba effect.
\cite{Nitta97,Engels97,Schaepers98b} The Rashba spin-orbit
coupling originates from a macroscopic electric field in an
asymmetric quantum well.\cite{Bychkov84} Meanwhile, research
activities have been extended to planar quasi one-dimensional
structures, which promise a superior spin
control.\cite{Nitta99,Kiselev01,Zuelicke02} The energy spectrum
and spin precession in these structures are governed by the
interplay between confinement and energy splitting due to
spin-orbit coupling. \cite{Guzenko06,Guzenko07} Only a few
theoretical investigations have dealt with the effect of
spin-orbit coupling in cylindrical conductors on the electronic
states and on the quantum
transport.\cite{Magarill98,Tserkovnyak06,Jin10,Entin-Wohlman10}
The spin-dynamics in curved two-dimensional electron gases was
discussed by Trushin and Schliemann \cite{Trushin08} while the
weak antilocalization effect in cylindrical wires was studied by
Wenk and Kettemann.\cite{Wenk10} The presence of spin-orbit
coupling was confirmed for InN \cite{Petersen09} and InAs
semiconductor nanowires by measuring the weak antilocalization
effect. \cite{Hansen05,Dhara09,Roulleau10,Estevez10}

The various possibilities of spin control in two-dimensional
electron gases and planar wire structures opened up by the Rashba
effect have inspired us to analyze theoretically the spin dynamics
in tubular conductors. We have used a cylindrical InAs nanowire
with a surface two-dimensional electron gas as a model system, but
our findings also apply to other systems, e.g. InN or InSb
nanowires. In Sect. \ref{Sec:II} we analyze the electronic states,
focusing on spin properties, and we discuss the conditions under
which a spin precession can be observed in tubular nanowires at
zero magnetic field (Sect.~\ref{Sec:III}) and in an axial magnetic
field (Sect.~\ref{Sec:IV}). In Sect.~\ref{Sec:V}, we comment on
the suitability of tubular conductors for spin electronic devices.

\section{Electrons in cylindrical wires} \label{Sec:II}
\begin{figure}[h]
\begin{center}
\includegraphics[width=1.0\columnwidth,angle=0]{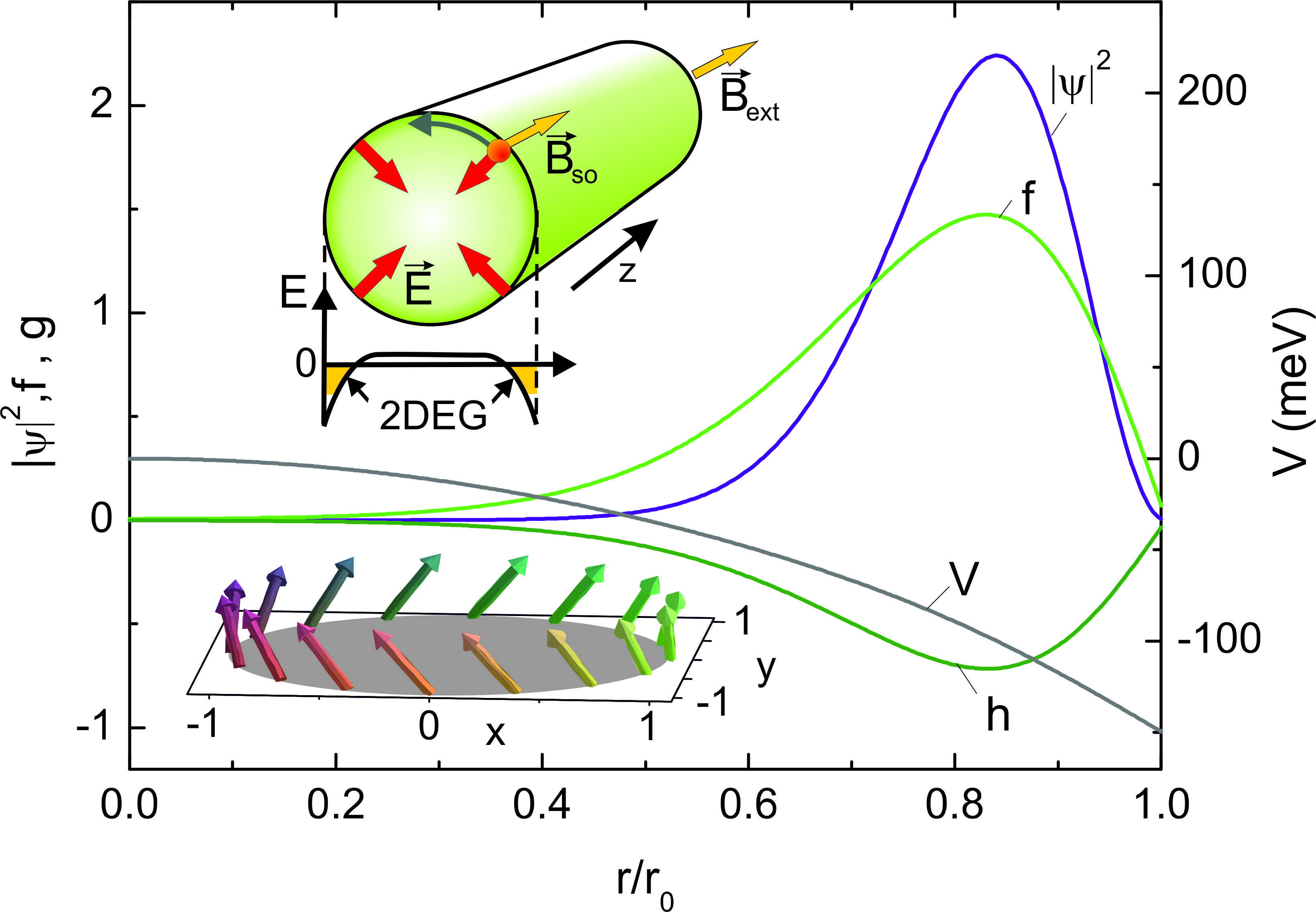}
\caption{Squared amplitude of the wave function $|\psi|^2$, the
spinor components $f$ and $h$ and potential profile $V$ as a
function of the normalized radius $r/r_0$. The upper inset shows a
schematic illustration of the nanowires, including the relevant
electric and magnetic fields. The lower inset shows the spin
orientation along the circumference for $j=1/2$. \label{fig:1}}
\end{center}
\end{figure}

Electrons confined in a cylinder move along the axis with a linear
momentum $\hbar k$ ($k$ real) and around the axis with an angular
momentum $\hbar l$ ($l$ integer). As long as the translational and
rotational symmetries of the cylinder are not perturbed these momenta
are conserved quantities.  The wave function of an electron
\[  \psi \ =\  \exp\left(\imath kz\right)
                   \exp\left(\imath l\phi\right)
                     f\left(r\right) \]
is a product of exponential functions in $z,\phi$, the coordinate
along the axis and the azimuthal angle around the axis
respectively, and a radial distribution function
$f\left(r\right)$. The distribution is determined by internal
forces produced by the cylinder material. In our case, we took a
planar 2-dimensional electron gas (2DEG) at the surface of  InAs
as a reference,\cite{Lamari03,Schierholz04} i.e. assuming a
surface state charge density of $N_S=1.27 \times
10^{11}$~cm$^{-2}$, a background $p$-doping of $n_d=2.8 \times
10^{17}$~cm$^{-3}$ and an effective electron mass of
$m^*=0.026\,m_e$. The calculations were done for a cylinder radius
$r_0=50$~nm. A schematic illustration of the nanowire is depicted
in Fig.~\ref{fig:1} (upper inset). Electrons of atoms at the surface
may find energetically more favorable states in the conduction band.
Due to the Coulomb attraction between the electrons and the ions
remaining at the surface the electrons get trapped in a layer close to
the surface forming a 2DEG.\cite{Smit89} The potential $V$
resulting from the charge density of occupied electron states
$\psi_{l,\sigma,k}$, of ions at the surface and of dopants $\rho_{BG}$
\begin{equation}
   \rho\ =\left(\ e\sum_{l,\sigma,k}^{occ}|\psi_{l,\sigma,k}|^2
                       \ +\ \rho_{BG}\ \right)/\epsilon_r
\label{dens}
\end{equation}
is shown in Fig.~\ref{fig:1}. $e$ is the elementary charge,
$\sigma$ the spin index. $\epsilon_r=14.6$ is the
bulk dielectric constant of InAs.\cite{Winkler03} It takes the polarization charges
of the medium into account. The potential profile is determined by
Poisson's equation which is solved in cylindrical symmetry
analytically
\begin{equation}
     V =\ 4\pi\epsilon_0 e
             \int_0^r
             r^{\prime}dr^{\prime}\rho\left(r^{\prime}\right)
                                          \ln\frac{r^{\prime}}{r}.
\label{vau}
\end{equation}
Equations~(\ref{dens}) and (\ref{vau}) are solved
self-consistently. Starting from the potential of a homogeneous
distribution of electrons in the cylinder the distribution is
recalculated using the Schr\"odinger equation given below [see
Eq.~(\ref{Schroe})] and Eq.~(\ref{dens}). The iteration procedure
converges monotonically. We assumed an interface barrier of
infinite height.

Due to the electric field $\vec{{\cal E}}=-\nabla V/e$ across the
surface of the cylinder the spin $\vec\sigma$ of the electron is
coupled to its orbital motion
\begin{eqnarray}
   {\cal H}_{SO} & = & \vec{\sigma}\cdot
       \left[\vec{p} \times e\vec{{\cal E}}\right]
              \frac{\gamma}{\hbar}   \nonumber  \\
                      & = &  \gamma V^{\prime}    \left[
        \left( \begin{array}{cc}
                         0      &    \imath\ e^{-\imath\phi}   \\
                     -\imath\ e^{\imath\phi}    &   0
                     \\
                  \end{array}    \right)
             \frac{\partial}{\imath\partial z} \right.  \nonumber
             \\
       & & \left.  + \left( \begin{array}{cc}
                                     1    &    0  \\
                                    0    &    -1           \\
                             \end{array}    \right)
                         \frac{\partial}{r\imath\partial\phi}
                         \right] \; .
\label{hso}
\end{eqnarray}
The coupling-strength $\gamma$ is determined by the band structure of
the cylinder material ($1.17$~nm$^2$ for InAs).\cite{Winkler03}
The second part of Eq.~(\ref{hso}) expresses ${\cal
H}_{SO}$ in terms of Pauli matrices for $\sigma_{x,y,z}$ acting on a
2-component (spinor) wave function $(\psi_{\uparrow},\psi_{\downarrow})
$. The off-diagonal terms in ${\cal H}_{SO}$ raise (lower) the value of
the orbital angular momentum ${\cal L}_z$ of $\psi_{\uparrow}$
($\psi_{\downarrow}$) by $\hbar$. The stationary states are eigenstates
of the total angular momentum
 ${\cal J}_z={\cal L}_z+{\cal S}_z$ (${\cal
 S}_z=\hbar\sigma_z/2$)
with eigenvalues $j=l\pm 1/2$. The spinor is of the form:
\begin{equation}
     \left( \begin{array}{c} \psi_{\uparrow} \\
                                 \psi_{\downarrow} \end{array}
                                 \right)
           = e^{\imath k z}e^{\imath l\phi}
               \left( \begin{array} {c} f\left(r\right) \\
                    \imath e^{\imath\phi}h\left(r\right)
                    \end{array}
                \right) \; ,
\end{equation}
where $f,h$ are real functions and solve the differential
equations
\begin{eqnarray}
            -\frac{\hbar^2}{2m^*}
         \left( f^{\prime\prime} + \frac{1}{r} f^{\prime}\right)
             + \left(\hat{V}_{l,+}-\hat{\epsilon} \right)\ f
       & = & k \gamma V^{\prime}\ h                  \nonumber\;,
       \\
            -\frac{\hbar^2}{2m^*}
         \left( h^{\prime\prime} + \frac{1}{r} h^{\prime}\right)
         + \left(\hat{V}_{l+1,-}-\hat{\epsilon} \right)\ h
         & = & k \gamma V^{\prime} \ f \; .  
\label{Schroe}
\end{eqnarray}
Here, $ \hat{V}_{l,\pm} = (\hbar l)^2/(2m^*r^2)+ V \pm\gamma
V^{\prime}l/r$ contains the contributions of the centrifugal force
and the diagonal spin-orbit term, $ \hat{\epsilon} = \epsilon -
(\hbar k)^2/(2m^*) $ is the energy without the axial kinetic
energy. At the wire boundary we assumed a barrier of infinite
height.\cite{foot01} The influence of an external magnetic field
$B$ is not included yet.
\begin{figure}[h]
\begin{center}
\includegraphics[width=1.0\columnwidth,angle=0]{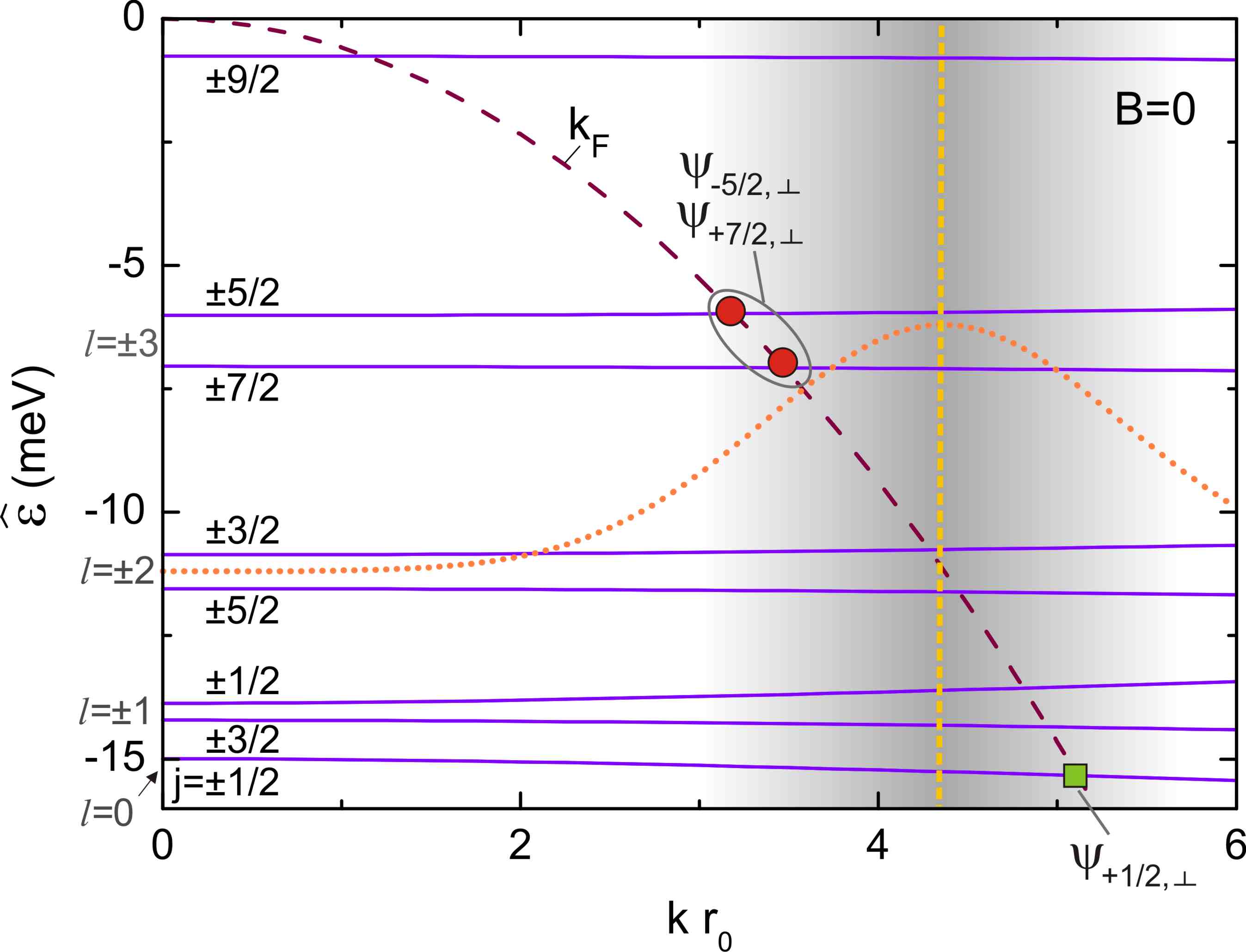}
\caption{Energy vs. $k$ dispersion at $B=0$. The dashed line
indicates the axial kinetic energy left out, which crosses the
bands at $k_F$. The pair of dots represent states forming the
superpositions states $\psi_{-5/2,\perp}$ and $\psi_{+7/2,\perp}$,
while the square indicates the state $\psi_{+1/2,\perp}$. The
dotted line illustrates the Gaussian wave packet of width
 $\delta k=1/r_0$.
\label{fig:2}}
\end{center}
\end{figure}

In Fig.~\ref{fig:2} the energy $\hat{\epsilon}$ is plotted for
several $j$-bands at $B=0$. The parabola indicates the axial
kinetic energy left out. It crosses the bands at the
Fermi-momentum $k_F$, i.e. states with energy below the parabola
are occupied.  At $k=0$ the coupling between $l$ and $l+1$
vanishes [cf. Eq. (\ref{hso}) ]. Classification with respect to
$l$ is possible. The splitting between the second and third band
($l=\pm 1$) is caused by the diagonal part of ${\cal H}_{SO}$ and
increases proportional to $l$ for the higher states. Due to the
mirror symmetry $z\leftrightarrow -z$ states with angular momentum
and spin reversed have the same energy. Therefore, all bands are
twofold degenerate.

\begin{figure}[]
\begin{center}
\includegraphics[width=1.0\columnwidth,angle=0]
                {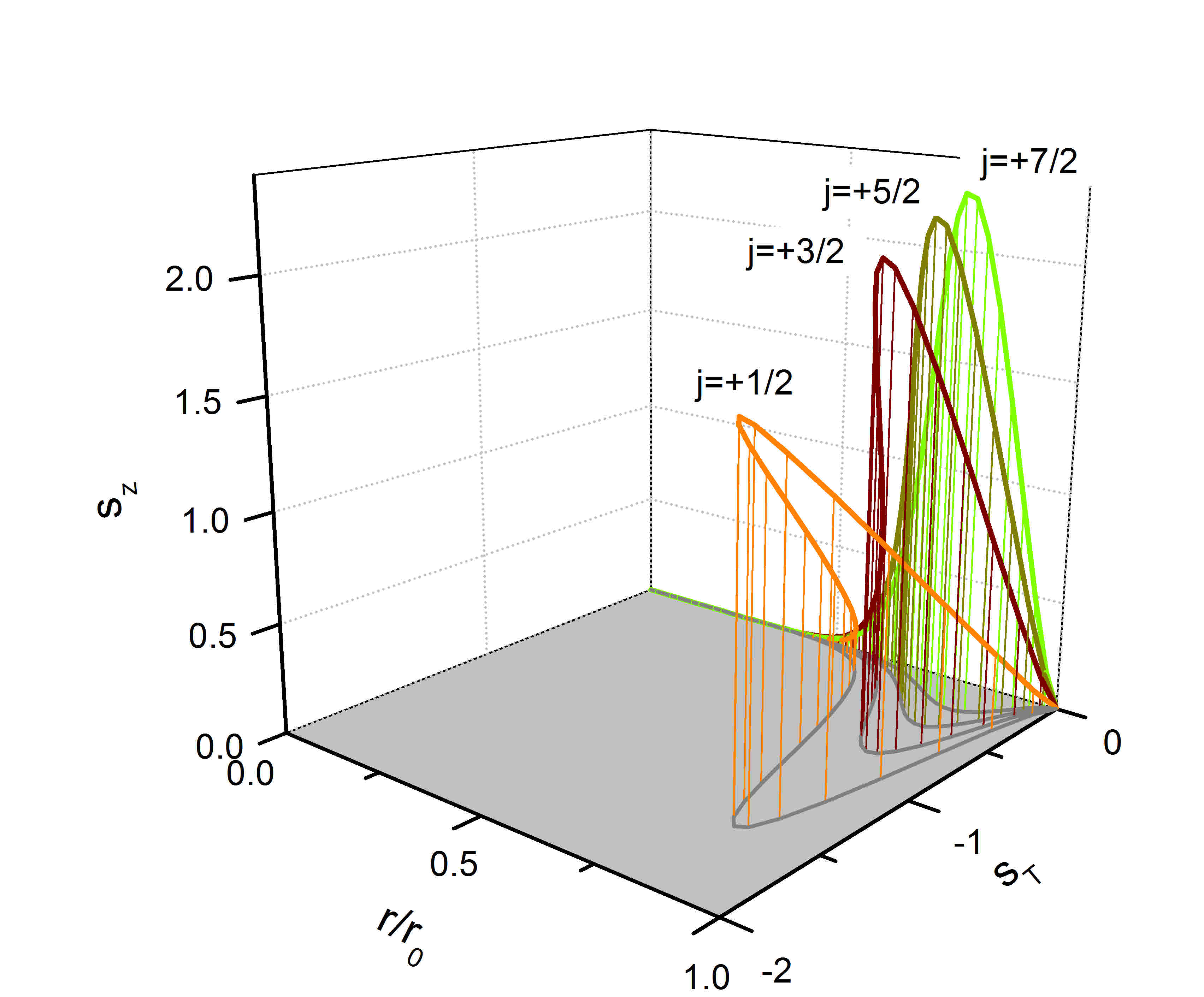}
\caption{Spin density $(s_T,s_z)$ of the lower energy states for
total angular momenta $j=1/2, 3/2, 5/2$ and $7/2$. The spin is
oriented only tangentially and along the $z$-axis. \label{fig:3}}
\end{center}
\end{figure}

The solution $(f,h)$ of Eq.~(\ref{Schroe}) for $j=1/2$ at $k_F$
is shown in Fig.~\ref{fig:1}. The spin-orbit coupling increases
linearly with $k$, i.e. at $k_F$ with $l=0$ there is the strongest
spin-orbit coupling. The spin density attains a sizable tangential
component
\begin{equation}
 s_T =   \left(
        \begin{array}{c} \psi_{\uparrow}^*   \\
                                \psi_{\downarrow}^* \end{array}
                     \right)\left(
        \begin{array}{cc}
                         0      &   -\imath\ e^{-\imath\phi}   \\
                     \imath\ e^{\imath\phi}    &   0           \\
         \end{array}    \right)
         \left( \begin{array}{c} \psi_{\uparrow} \\
                                 \psi_{\downarrow} \end{array}
                                 \right)
                = 2fh.
\nonumber
\end{equation}
The component along the wire axis is
\begin{equation}
 s_z  =   \left(
        \begin{array}{c} \psi_{\uparrow}^* \\
                                \psi_{\downarrow}^* \end{array}
                     \right)\left(
        \begin{array}{cc}
                         1      &    0   \\
                         0     &    -1   \\
         \end{array}    \right)
         \left( \begin{array}{c} \psi_{\uparrow} \\
                                 \psi_{\downarrow} \end{array}
                                 \right)
                = f^2-h^2.
\label{spidi}
\end{equation}
The radial component is zero. The spin orientation around the
cylinder for $j=1/2$ is illustrated in Fig.~\ref{fig:1} (lower
inset). According to Eq.~(\ref{spidi}) the spin turns to the axial
direction. This is shown for different values of $j$ in the plot
of the spin densities $s_T$ and $s_z$ in Fig.~\ref{fig:3}. As can
be seen here, the spin is oriented exclusively tangentially and
along the axial direction. When averaged over the cylinder plane
$\langle\cdots\rangle$ for each state $\psi_j$ the spin components
$\langle\sigma_x\rangle$ and $\langle\sigma_y\rangle$ are zero,
while a finite contribution $\langle\sigma_z\rangle$ remains along
the $z$-direction.

\section{Superposition States and Spin Precession}
\label{Sec:III}

For each $k$ and $j$ there are two solutions of Eq.~(\ref{Schroe})
$\psi_{j,\pm}$. The (+)-state originates from $(\psi_l,0)$, the
($-$)-state from $(0,\psi_{l+1})$, the solutions at $k=0$, with
$j=l+1/2$. They are orthogonal to each other and have opposite spin
direction ($\pm$). They have different energies $\hat{\epsilon}$ and
therefore different $k_F$. Their superposition
$\psi_{j,\parallel}=(\psi_{j,+}+\psi_{j,-})/\sqrt{2}$ yields
\begin{eqnarray}
 \langle\sigma_z\rangle_\| &=& \left(
                   \langle f_{j,+}^2-h_{j,+}^2\rangle
                + \langle f_{j,-}^2 - h_{j,-}^2\rangle \right)/2
  \nonumber \\
               &+& \langle f_{j,+}f_{j,-}-h_{j,+}h_{j,-}\rangle
                     \cos\left( k_{F,+}-k_{F,-}\right)z \;.
                     \nonumber
\end{eqnarray}
The contributions of the basis states ($\pm$) almost cancel each
other and are neglected further on. The interference between the
states is constructive due to orthogonality and leads to
\begin{equation}
 \langle\sigma_z\rangle_\| \approx 2\langle f_{j,+}f_{j,-}\rangle
                      \cos\left( k_{F,+}-k_{F,-}\right)z \; ,
\label{SpiWav}
\end{equation}
an oscillation of the average spin along the cylinder axis with a
wavelength $\lambda_{\parallel} = 2\pi/|k_{F,+}-k_{F,-}|$. The
spin components $\langle\sigma_x\rangle_\|$,
$\langle\sigma_y\rangle_\|$ in the cylinder plane are both zero.

Superpositions of eigenstates with different $j$'s form states with a
non-zero average spin component in the cylinder plane, e.g.
$\psi_{j,\perp} = (\psi_{j,+}+\psi_{j-1,-})/\sqrt{2}$. As one can
easily retrace, these states originate from states with the same
angular momentum $l$. The interference term gives the only
$\phi$-independent contribution to the densities of
$\sigma_x,\sigma_y$. With
\begin{eqnarray}
 \psi_{j,+} &=&  \exp\left(\imath zk_{F,+}\right)
                   \exp\left(\imath l\phi\right)
               \left( \begin{array} {c}f\left(r\right) \\
                    \imath e^{\imath\phi}h\left(r\right)
                         \end{array} \right)\; , \nonumber \\
                          \psi_{j-1,-} & =&  \exp\left(\imath
z\tilde{k}_{F,-}\right)
                   \exp\left[\imath \left( l-1 \right)\phi\right]
               \left( \begin{array} {c} \tilde{f}\left(r\right) \\
                    \imath e^{\imath\phi}\tilde{h}\left(r\right)
                    \end{array}
                \right) \; , \nonumber \\
                         \end{eqnarray}
the averages are
\begin{eqnarray}
 \langle\sigma_x\rangle_\perp &=&  \langle h\tilde{f} \rangle
      \sin\left(k_{F,+}-\tilde{k}_{F,-}\right)z \; , \nonumber \\
 \langle\sigma_y\rangle_\perp &=& \langle h\tilde{f} \rangle
       \cos \left(k_{F,+}-\tilde{k}_{F,-}\right)z \; .
\label{SpiPrae}
\end{eqnarray}
The $\langle\sigma_z\rangle_\perp$ contribution is small and does
not depend on $z$. In particular, for the superposition
$\psi_{1/2,\perp}$ of the lowest two states
$\langle\sigma_z\rangle_\perp$ is zero.

\begin{figure}[h!]
\begin{center}
\includegraphics[width=0.40\columnwidth,angle=0]
                {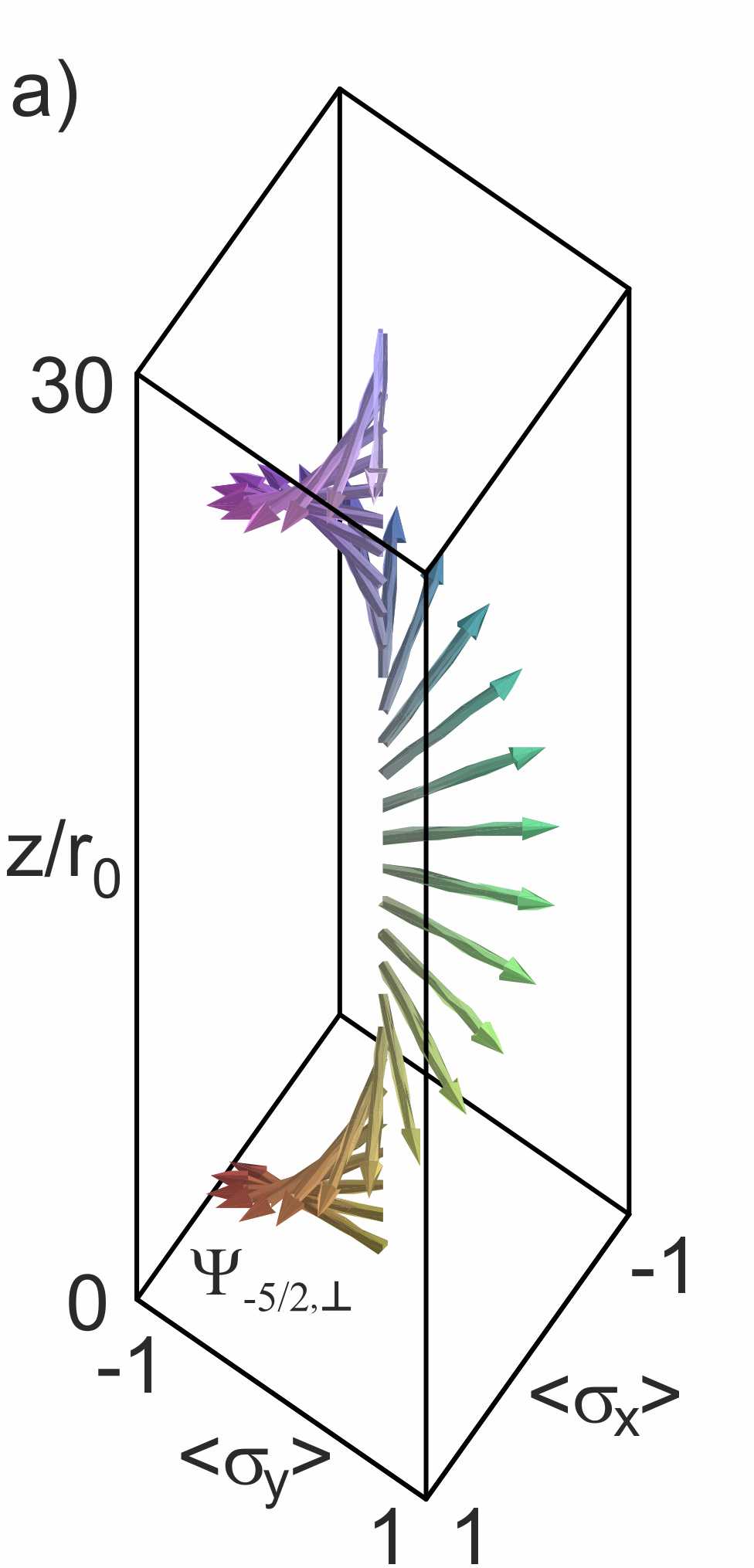}
\includegraphics[width=0.40\columnwidth,angle=0]
                {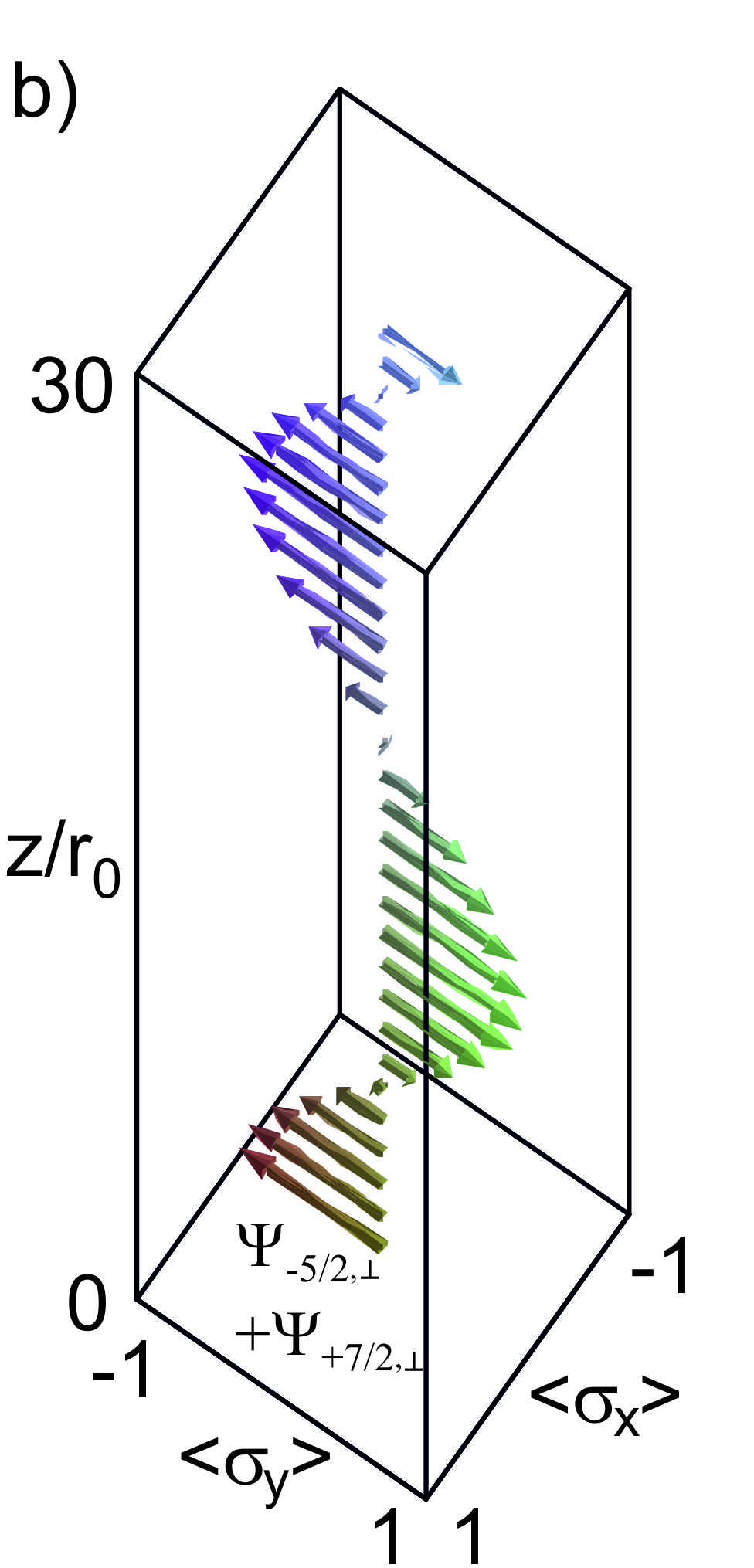}
\includegraphics[width=0.40\columnwidth,angle=0]
                {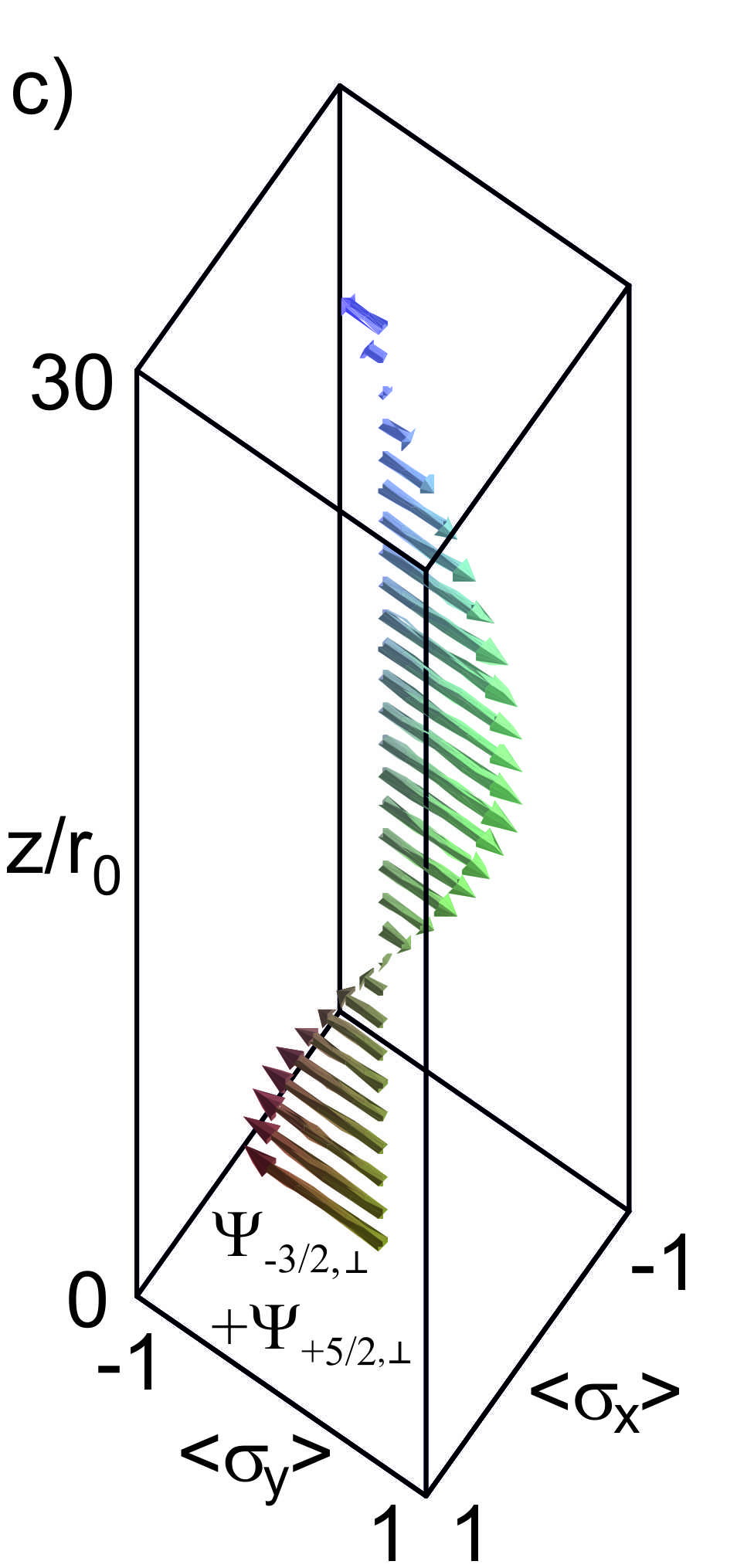}
\includegraphics[width=0.40\columnwidth,angle=0]
                {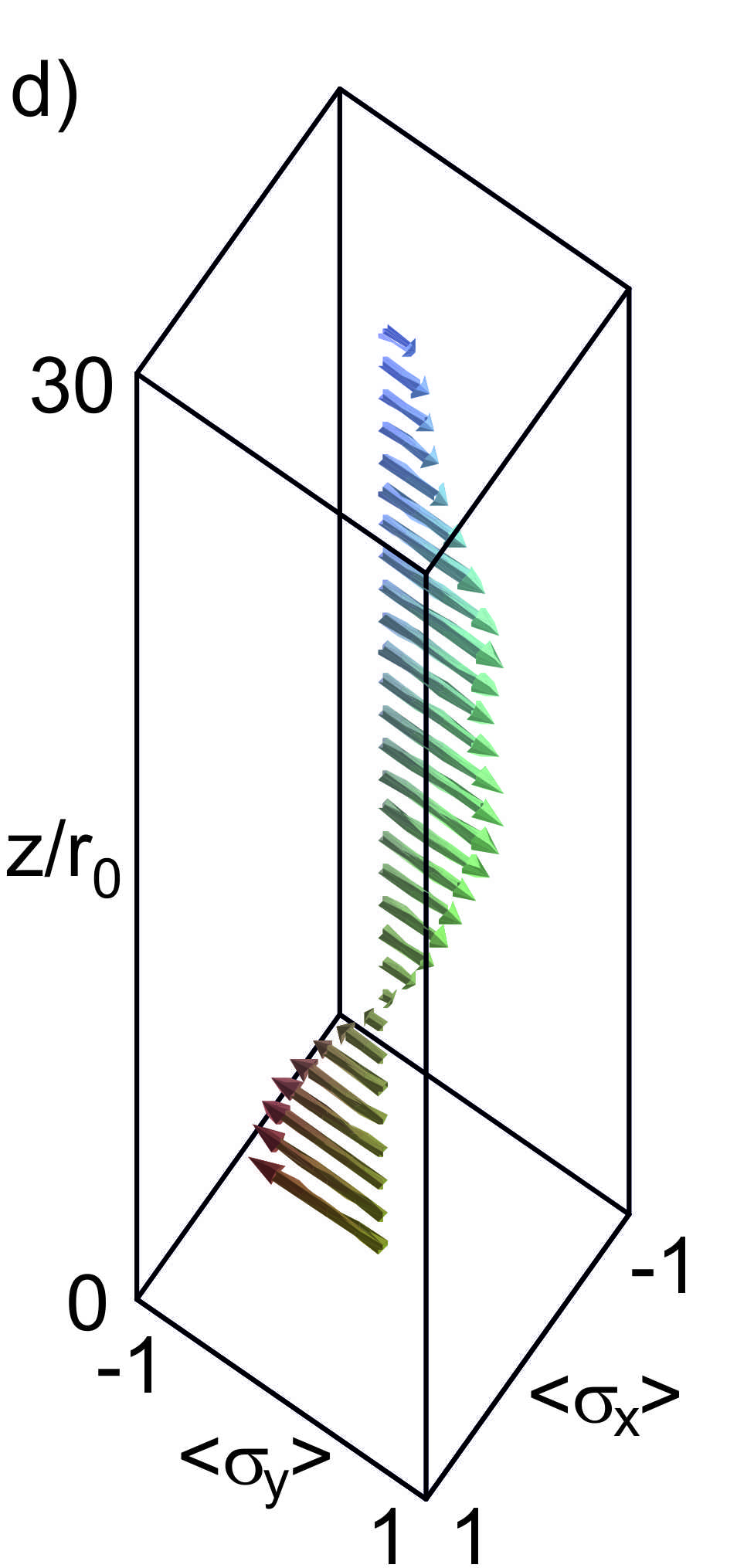}
\caption{(a) Counter-clockwise spin precession of electrons in the
superposition state  $\psi_{-5/2,\perp}$ at the Fermi energy
constituted of the states $\psi_{-5/2,+}$ and $\psi_{-7/2,-}$ for
a propagation along the wire axis from $z/r_0=0$ to $30$. (b) Spin
orientation of the sum of the contribution shown in (a) and the
corresponding clockwise contribution $\psi_{+7/2,\perp}$ being a
superposition of $\psi_{+7/2,+}$ and $\psi_{+5/2,-}$. (c) Spin
oscillations resulting from the combinations of the two lower
energy superposition states $\psi_{-3/2,\perp}$ and
$\psi_{+5/2,\perp}$. (d) Spin variation for a \mbox{Gaussian}
wave-packet of width $1/r_0$ centered between the $k_F$'s of the
states $\psi_{-3/2,\perp}$ and $\psi_{+5/2,\perp}$ (cf.
Fig.~\ref{fig:2}).
 \label{fig:4}}
\end{center}
\end{figure}

For  $\psi_{-5/2,\perp}$, the superposition of $\psi_{-5/2,+}$ and
$\psi_{-7/2,-}$, the spin precesses counterclockwise in the
cylinder plane along the cylinder axis, as illustrated in
Fig.~\ref{fig:4}(a).
Here, we assumed an initial spin orientation along the $-y$
direction, which in practice can be realized by spin injection
from a spin-polarized electrode. For $\psi_{+7/2,\perp}$
constituted of the opposite states $\psi_{+7/2,+}$ and
$\psi_{+5/2,-}$ the spin precession is clockwise. Both precessions
have the same period of
$\lambda_{\perp}=2\pi/|k_{F,+}-\tilde{k}_{F,-}|$. Their energy is
degenerate. Due to their exactly inverse precession sense the
combination of these states results in an oscillatory behavior of
the net spin orientation, as depicted in Fig.~\ref{fig:4}(b). For
an initial spin orientation along the $-y$ direction the spin
oscillates in the $yz$-plane. Superposition of the respective
opposite states restores the left-right symmetry and eliminates
spin precession. The oscillation period $\lambda_{\perp}$ of
$\psi_{j,\perp}$ depends on $j$. For smaller $|j|$, e.g.
$\psi_{-3/2,\perp}$ the corresponding difference in $k_{F,+}$ and
$\tilde{k}_{F,-}$ becomes smaller so that the period
$\lambda_{\perp}$ is enlarged, as one can infer from
Fig.~\ref{fig:4}(c) as compared to Fig.~\ref{fig:4}(b). The
superposition state $\psi_{+1/2,\perp}$ constituted of the two
lowest lying energy states $\psi_{\pm 1/2,\pm}$ (cf.
Fig.~\ref{fig:2}, square) shows no precession at all, because here
$k_{F,+}$ and $\tilde{k}_{F,-}$ are identical.
Figure~\ref{fig:4}(d) shows the spin variation for a Gaussian wave
packet of width $\delta k=1/r_0$ centered between the $k_F$'s of
the states $\psi_{-3/2,\perp}$ and $\psi_{+5/2,\perp}$. In
position space this corresponds to a distribution of width $2r_0$.
The oscillation deviates from a purely harmonic oscillation, as
shown in Fig.~\ref{fig:4}(c), due to the contributions of the
other states at the Fermi energy. This effect is also increasing
with decreasing $|j|$ when the $k_F$'s get closer to each other.

The electron spin is usually injected from a spin-polarized electrode
in all states at the Fermi energy $E_F$ having the correct spin
direction. Thus, if only the direction of the spin is fixed by the
electrode, all states are likely to transport electrons through the
cylinder and a definite precession will not be observed. The total spin
will only vary in the plane which is defined by the initial spin
orientation and the $z$-axis, similar to the situation illustrated in
Fig.~\ref{fig:4}. In order to observe spin precession about the
cylinder axis, a selection mechanism which breaks the left-right
symmetry of the system must be adopted. As it will be discussed in the
next section, this is achieved by applying a longitudinal magnetic
field $\vec{{\cal B}}=(0,0,B)$.

\section{Spin Precession in a Magnetic Field} \label{Sec:IV}

\begin{figure}
\begin{center}
\includegraphics[width=0.9\columnwidth,angle=0]{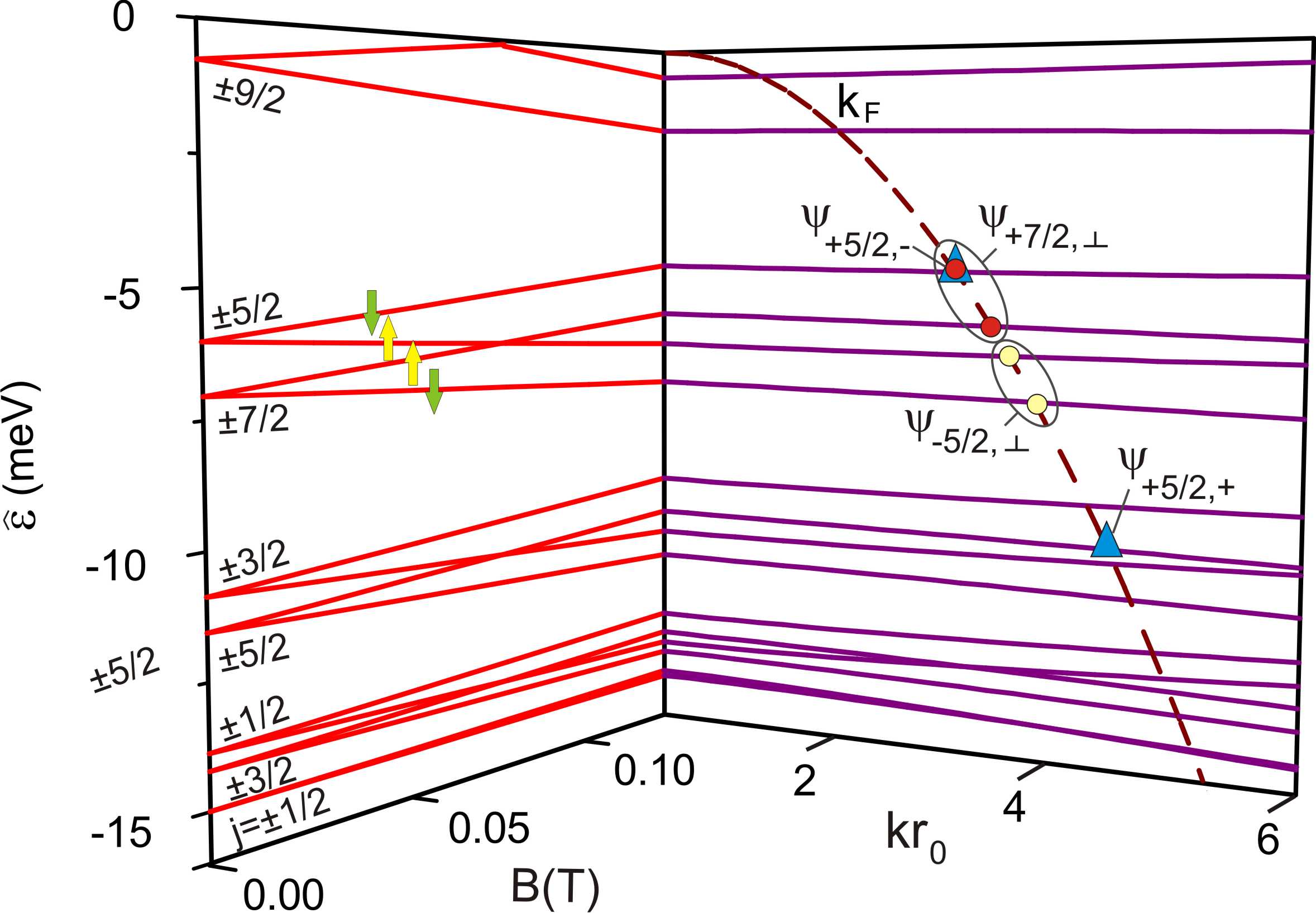}
\caption{Energy vs. $B$ dispersion (left panel) at $k=0$ and
energy vs. $k$ dispersion (right panel) at $B=0.13$~T. The dashed
line indicates the axial kinetic energy left out, which crosses
the bands at $k_F$. The pairs of dots indicate the states forming
the superposition states $\psi_{-5/2,\perp}$ and
$\psi_{+7/2,\perp}$ at $k_F$ with a net spin in the cylinder
plane. The two states $\psi_{+5/2,+}$ and $\psi_{+5/2,-}$ with
$j=+5/2$ are marked by triangles. \label{fig:5}}
\end{center}
\end{figure}

The vector potential $\vec{{\cal A}}=(-By/2,Bx/2,0)$ of a
longitudinal
magnetic field introduces a paramagnetic (Zeeman-) and
diamagnetic
(Landau-) term into Eq.~(\ref{Schroe}).
 $\hat{V}_{l,\pm}$ is extended to
$$ \tilde{V}_{l,\pm}
       = \hat{V}_{l,\pm} + \frac{\hbar e}{2m^*} B
                              \left(l\pm \frac{gm^*}{2m_e}\right)
          + \frac{e^2B^2}{8m^*} r^2 \; ,$$
with $g$ the gyromagnetic-factor of the electron spin ($-14.9$ for
InAs\cite{Winkler03}). The paramagnetic (second) term in
$\tilde{V}_{l,\pm}$ raises $\hat{\epsilon}$ for states with $j({\rm
or}\ l)>0$ and lowers $\hat{\epsilon}$ for states with $j({\rm or}\
l)<0$. The energy difference increases $\propto lB$ for $B\ll
l\hbar/(er_0^2)$ (cf. Fig.~\ref{fig:5}). For larger $B$
$\hat{\epsilon}$ increase $\propto B^2$ due to the diamagnetic (third)
term. In the linear range the influence of the $r$-dependence of the
third term is negligible. The densities do not change significantly.

\begin{figure}[]
\begin{center}
\includegraphics[width=0.40\columnwidth,angle=0]{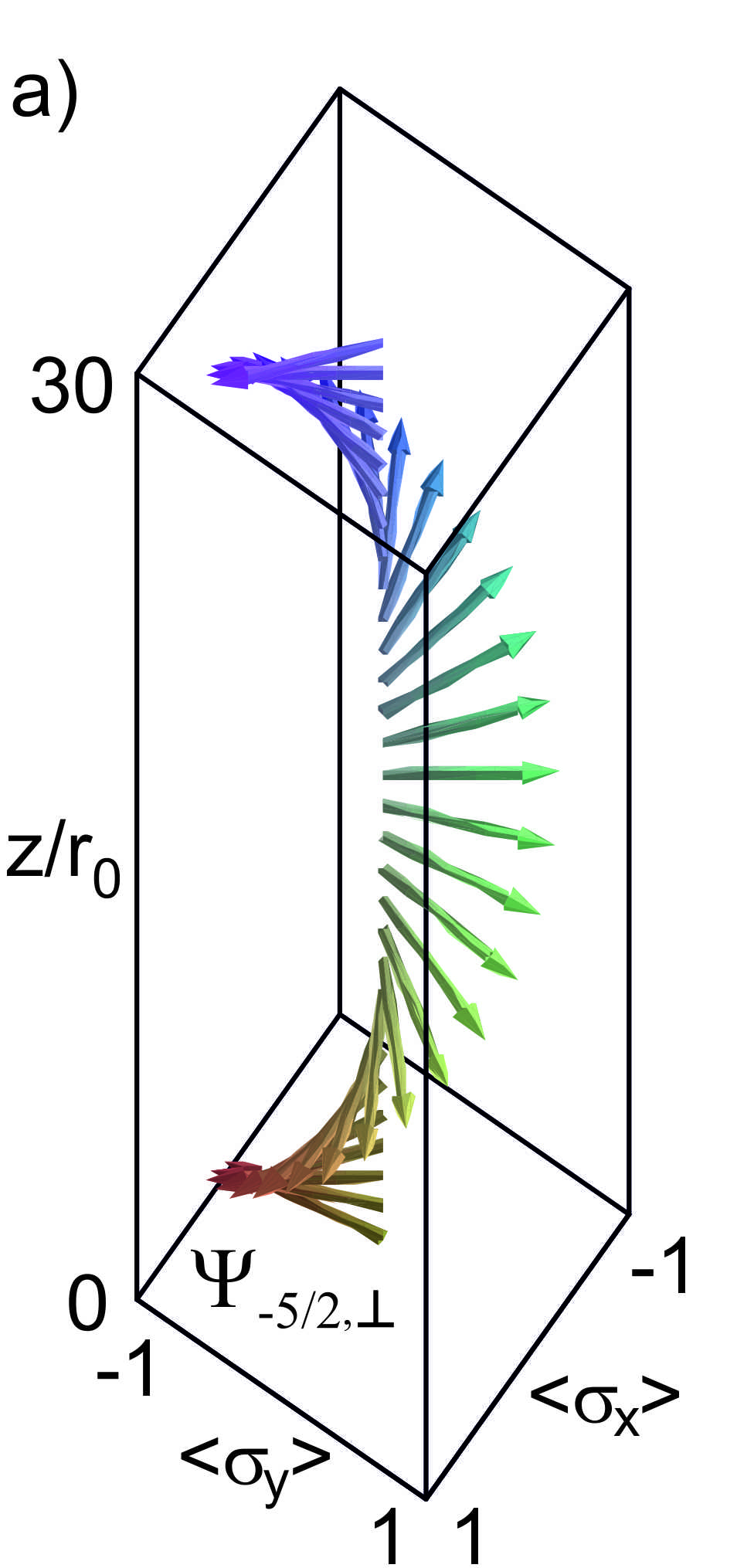}
\includegraphics[width=0.40\columnwidth,angle=0]{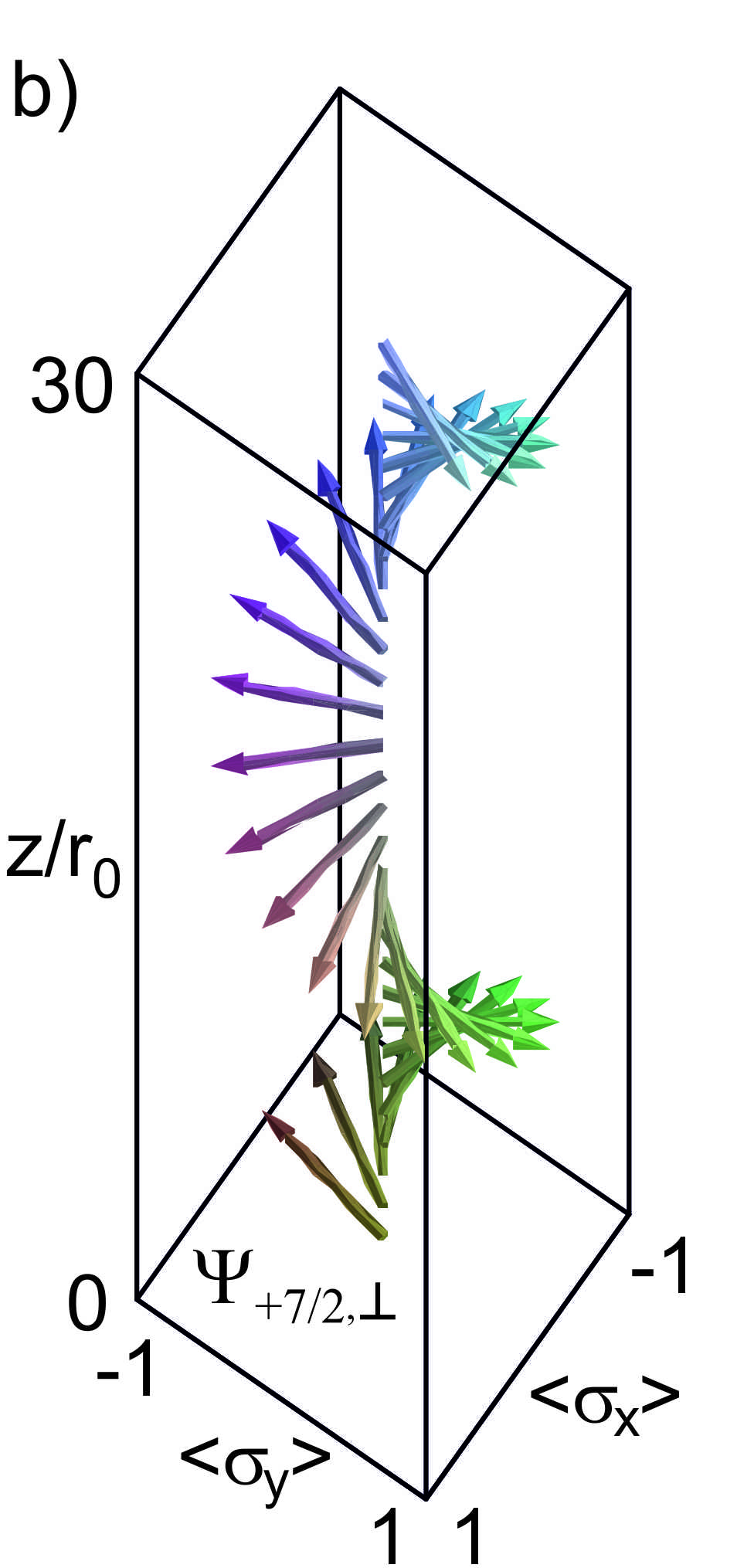}
\includegraphics[width=0.80\columnwidth,angle=0]{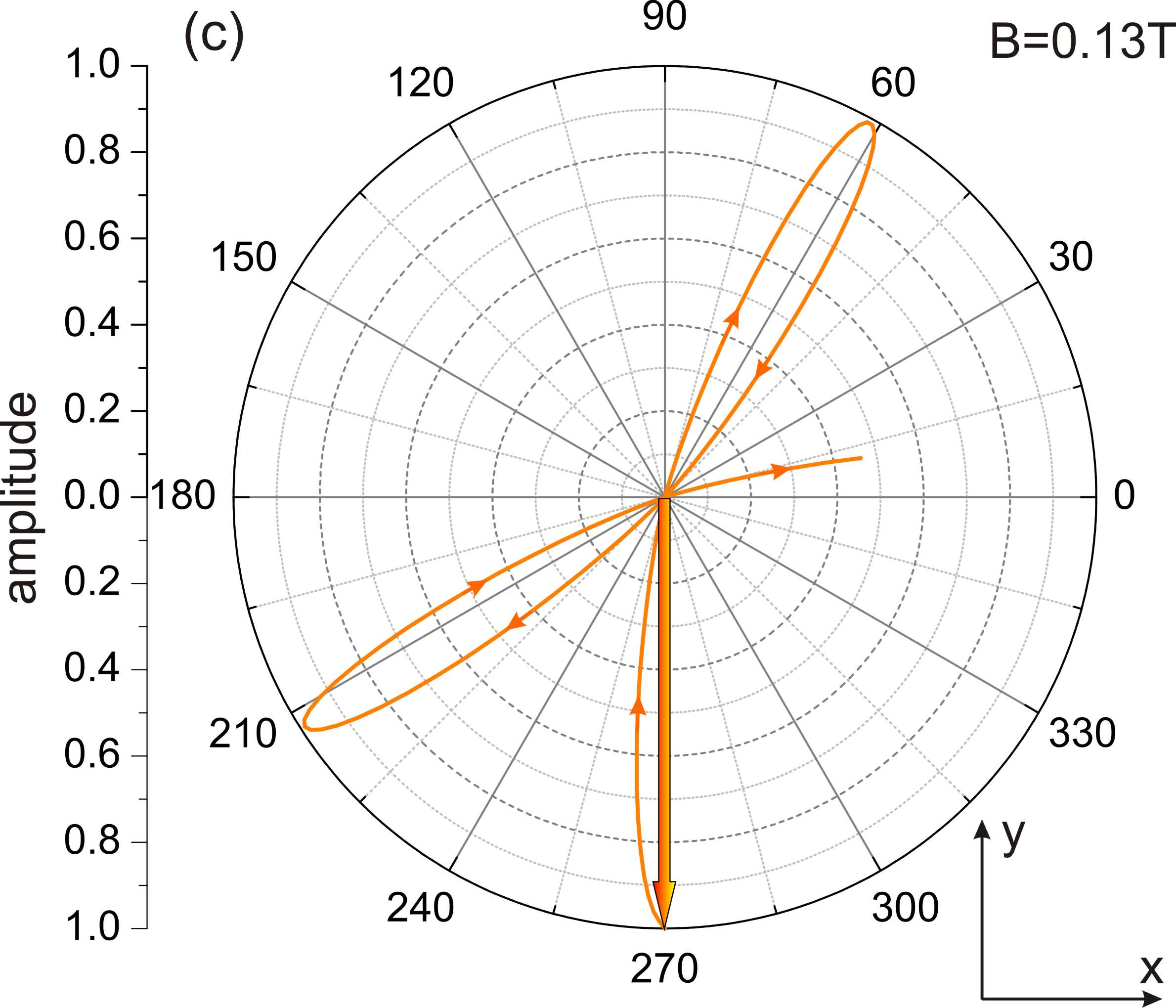}
\caption{(a) Spin precession of electrons at the Fermi energy
propagating along the wire axis for the superposition state
$\psi_{-5/2,\perp}$ at $B=0.13$~T. (b) Corresponding spin
precession for the state $\psi_{+7/2,\perp}$. (c) Spin orientation
and magnitude of the sum of the contributions shown in (a) and (b)
for a propagation from $z/r_0=0$ to $30$. The arrow indicates the
direction of the initially injected spin. \label{fig:6}}
\end{center}
\end{figure}

The main effect of $B$ is the energetic separation of the $\pm
j$-states. It opens possibilities of observing spin dynamics in
electronic transport. This will be demonstrated in the following
at $B=0.13$~T.  Figure~\ref{fig:5} shows the $B$-dependence at
$k=0$ up to $B=0.13$~T and the $k$-dependence at $B=0.13$~T of
$\hat{\epsilon}$ for states from $j=\pm 1/2$ to $\pm 9/2$. Again,
the parabola marks the Fermi edge. Superpositions with spin in the
cylinder plane according to Eq.~(\ref{SpiPrae}), $\psi_{j,\perp}$
are marked as pairs in Fig.~\ref{fig:5}. The lower pair
corresponds to $\psi_{-5/2,\perp}$ depicted in
Fig.~\ref{fig:4}(a). As illustrated in Fig.~\ref{fig:6}(a), it
shows the same counter-clockwise precession.
In contrast to the zero field case, now the superposition state
$\psi_{+7/2,\perp}$ has a larger $k_F$-difference, i.e. a shorter
precession length [cf. Fig.~\ref{fig:6}(b)]. Consequently, the
precessions of $\psi_{-5/2,\perp}$ and $\psi_{+7/2,\perp}$ are not
exactly opposite. In contrast to the case at $B=0$, the spin still
rotates following the state with the faster precession, when both
states are superposed. This is illustrated in Fig.~\ref{fig:6}(c),
where one finds that in addition to the oscillation of the spin
amplitude its orientation is also changed during propagation.
Thus, by applying a magnetic field a spin precession can be
achieved.
\begin{figure}[]
\begin{center}
\includegraphics[width=1.0\columnwidth,angle=0]{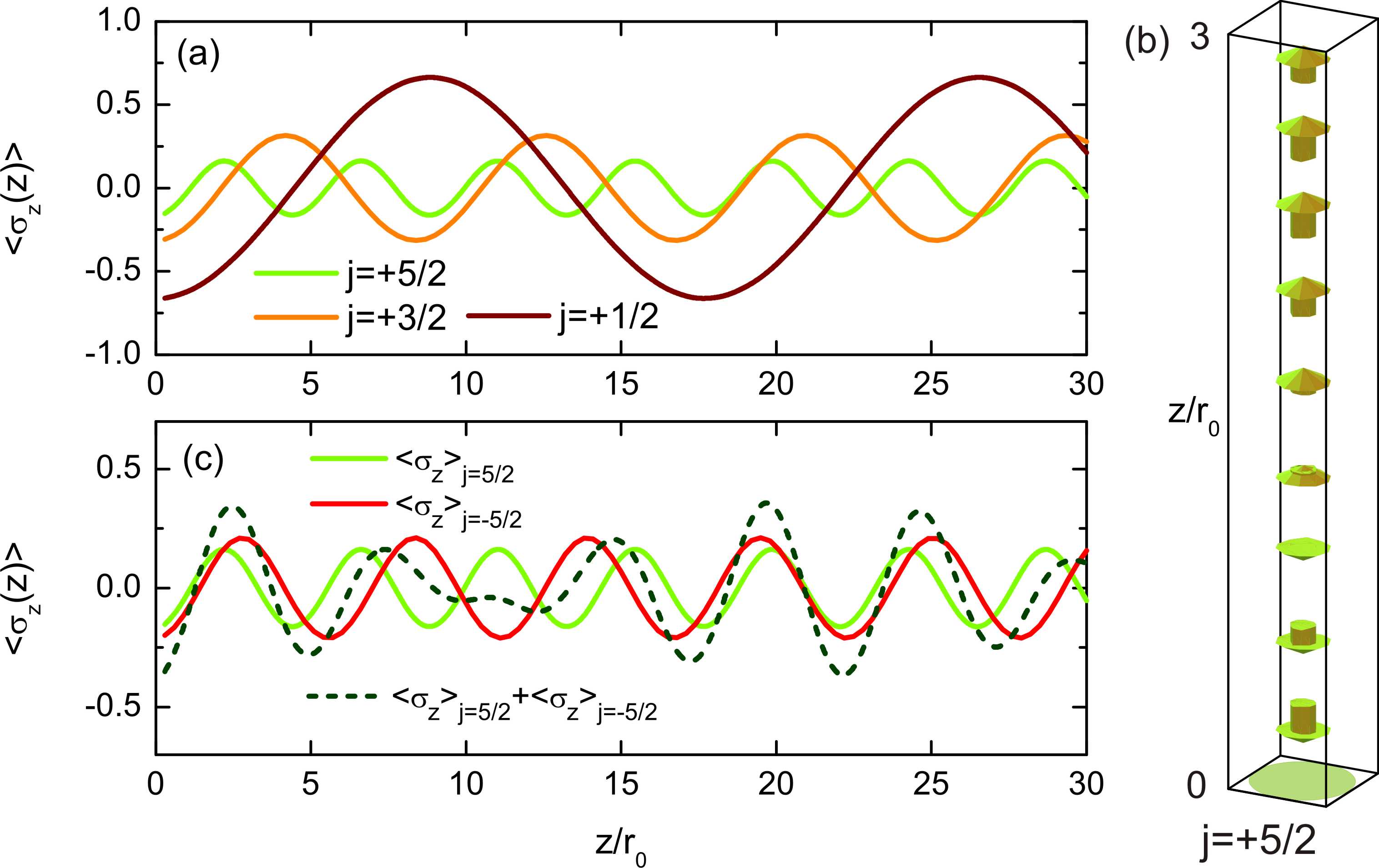}
\caption{(a) Spin orientation $\langle\sigma_z\rangle$ along the
wire axis for the superposition states $\psi_{j,\parallel}$ with
$j=+1/2, +3/2$ and $+5/2$ at $B=0.13$~T. (b) Illustration of the
oscillation of the average spin along the wire axis for the
superposition state $\psi_{+5/2,\parallel}$. (c) Modulation of the
total spin orientation $\langle\sigma_z\rangle$ (dashed line)
resulting from a combination of the $\psi_{-5/2,\parallel}$ and
$\psi_{+5/2,\parallel}$ states at $B=0.13$~T. \label{fig:7}}
\end{center}
\end{figure}

In the previous section, we already pointed out that the
superposition state $\psi_{j,\parallel}$ with equal total angular
momentum but opposite spin orientation result in an oscillation of
the average spin along the cylinder axis. In Fig.~\ref{fig:7}(a)
and (b) these oscillations of $\langle\sigma_z\rangle_\|$ are
shown for different values of $j$ at $B=0.13$~T. One finds that
for larger total angular momentum values the oscillation period is
shorter owing to the larger difference of Fermi wave vectors. In
Fig.~\ref{fig:5} the states contributing to
$\psi_{+5/2,\parallel}$ are marked by triangles. Compared to the
previously discussed $\psi_{j,\perp}$ states, here the difference
in the Fermi vectors is relatively large, leading to a faster
oscillation compared to the spin precession period shown in
Fig.~\ref{fig:6}(b).

Once again the application of an axial magnetic field breaks the
symmetry of the $\psi_{\pm j,\parallel}$ states. As can be inferred
from Fig.~\ref{fig:7}(c), a different oscillation period is found for
the $\psi_{+5/2,\parallel}$ and $\psi_{-5/2,\parallel}$ states. Thus,
when these states are combined a beating in the oscillation of the
average spin appears.

\section{Conclusions} \label{Sec:V}

In the previous two sections we learned that an injected spin is
strongly modulated while propagating through a cylindrical
nanowire. For a spin injection along the wire axis, e.g. by a
ferromagnetic electrode, the spin is carried by superposition
states with equal total angular momenta. In analogy to the spin
field-effect transistor based on a planar 2DEG,\cite{Datta90} a
transistor structure can be realized by placing a second magnetic
electrode at the opposite terminal of the nanowire as a spin
detector. Control of the spin orientation can achieved by
manipulating the strength of the Rashba effect by means of a gate
electrode. By applying a bias voltage to the gate, the strength of
the electric field $\vec{\mathcal{E}}$ in the surface 2DEG is
adjusted. In order to obtain a uniform control within the channel,
a so-called wrap-around gate should be preferred.\cite{Bryllert06}
Usually, in a realistic situation a larger number of states with
different total angular momenta $j$ is occupied. As we observed,
for each superposition state $\psi_{j,\parallel}$ different
oscillation periods are found. This leads to a rather complex
modulation of the spin along the axial direction. An obvious
strategy for simplification is to reduce the number of occupied
states, i.e. by depleting the channel by means of a gate. Another
possibility might be to only occupy certain states by means of
$k$-selective filters. This might be realized by means of an
injection through a single or a resonant tunneling barrier. As
pointed out in Sect.~\ref{Sec:III}, one possible way to model this
situation is to assume the formation of a state with a Gaussian
distribution around the average momentum.

In addition to a spin injection and detection along the wire axis it is
also possible to inject spins in transversal direction. Here, the spins
are carried by superposition states $\psi_{j,\perp}$ constituted of
states with different total angular momenta $j$. As long as no magnetic
field is applied the spin is exclusively modulated in the plane spanned
by the injection orientation and the wire axis. Here, the output signal
in a spin field-effect transistor is gained by gate-modulating the spin
orientation along or opposite to a detector electrode, which is
polarized parallel or antiparallel to the injector. By applying an
axially oriented magnetic field, spin precession about the wire axis
can be achieved. This additional feature might be an interesting option
to implement more complex functionalities in spin electronic devices.

In conclusion, we have shown that semiconductor nanowires affected by
Rashba spin-orbit coupling are promising candidates for future
nanowire-based spin electronic devices. The complex spin dynamics in
these cylindrically-shaped conductors provide many opportunities to
tailor the device functionality.

\acknowledgements We thank N. Demarina (Forschungszentrum
J\"ulich) for fruitful discussions regarding the
Schr\"odinger-Poisson solver in cylindrical systems and U.
Z\"ulicke (Massey University, New Zealand) and R. Winkler
(Northern Illinois University, USA) on the Rashba effect at
semiconductor interfaces. Furthermore, we acknowledge the support
of or work by S. Bl\"ugel and D. Gr\"utzmacher (Forschungszentrum
J\"ulich). This work was supported by the Deutsche
Forschungsgemeinschaft through FOR 912.


\begin{thebibliography}{36}
\bibitem[{foo()}]{foot01}
\bibinfo{note}{We are aware that by our approach of assuming an infinite
  barrier at the wire boundary the leakage of the wavefunction into the barrier
  is neglected. It has been shown that the leakage of the wavefunction can
  contribute significantly to the Rashba effect.\cite{Schaepers98b,Winkler03}
  However, since the InAs interface to the environment, e.g. to vacuum, cannot
  be described with sufficient accuracy, we decided to assumed a barrier of
  infinite height and adjusted the strength of spin-orbit coupling to the
  experimental data.\cite{Schierholz04}}.
\end{thebibliography}

\end{document}